\documentclass[
    ,final            
  ]
  {aipproc}

\layoutstyle{6x9}

\newcommand{\ffake}{f_{\xi<5\times 10^{-6}}}

\begin{document}

\title[The ATLAS Inelastic Cross-Section Measurement]{Measurement of the Inelastic Cross-Section and Prospects for Elastic Measurements with ATLAS}

\author{Lauren Tompkins$^*$ on behalf of the ATLAS Collaboration\footnote{Copyright CERN for the benefit of the ATLAS collaboration.}}{
  address={$^*$Lawrence Berkeley National Laboratory, 1 Cyclotron Rd. Berkeley, CA 94720, USA}
}

\begin{abstract}
A first measurement of the inelastic cross-section of proton-proton
collisions at $\sqrt{s}~=$~7~TeV using the ATLAS detector at the Large Hadron
Collider is presented. The measurement is made using scintillators in the
forward region of the ATLAS detector.  Prospects for elastic cross-section measurements are also discussed.
\end{abstract}

\maketitle


\section{Introduction}

The measurement of the proton-proton inelastic cross-section represents a benchmark in the opening of a new hadron collider energy frontier. The cross-sections can not yet be calculated by quantum chromodynamics, and many approaches
have been used to describe the existing data.  Therefore, experimental measurements are critical for developing a better understanding of these interactions.

Proton-proton interactions are divided into two categories for convenience: diffractive and non-diffractive events.  In diffractive events the protons interact via colorless object exchange, leading to large rapidity gaps between the proton dissociation products.  In non-diffractive events, colored objects are exchanged, resulting in a small probability for significant rapidity gaps.   The detector is insensitive to events with large rapidity gaps, so the measurement is quoted for the limited range of interactions satisfying $\xi = M_X^2/s > 5 \times 10^{-6}$, where $X$ is the largest mass diffractive system in the event and $\sqrt{s}$ is the center-of-mass energy. The measurement is additionally extrapolated to the full inelastic cross-section using model-dependent extrapolation factors.  

  These proceedings report on the first measurement of the proton-proton inelastic cross-section, the details of which are contained in~\cite{bib:mypaper}.  They additionally comment on the future prospects for measuring the elastic proton-proton cross-section.
\section{Event Selection and Acceptance}

The data were acquired using the ATLAS detector~\cite{bib:detpaper} during an early LHC run at $\sqrt{s} = 7$ TeV in $2010$, corresponding to an integrated luminosity of approximately $20~\mu$b$^{-1}$.  The luminosity was measured using a forward Cherenkov light detector, LUCID, calibrated to a precision of 3.4\% using dedicated beam separation scans~\cite{bib:lumiconf}.  

Events were recorded if at least one scintillator counter passed an online charge discriminator trigger.  The scintillators, collectively referred to as the Minimum Bias Trigger Scintillators (MBTS), consist of two wheels located at $z=\pm$3.6~m along the beam-pipe from the nominal interaction point (IP)\footnote{ATLAS uses a right-handed coordinate system centered at the IP with the $z$-axis aligned along the LHC beam pipe. The pseudorapidity is defined in terms of the polar angle $\theta$ as $\eta=-\ln\tan(\theta/2)$. }.  Each wheel is divided into two rings in $\eta$ and eight sections in $\phi$, for a total of 32 independent counters.  The offline selection required charge depositions in at least two forward scintillators, and the trigger requirement was found to be 99.98$^{+0.02}_{-0.12}$\% efficient with respect to this selection. 

The Monte Carlo models {\sc Pythia 6}~\cite{bib:py6}, {\sc Pythia 8}~\cite{bib:py8} and {\sc Phojet}~\cite{bib:pho1} were used to predict properties of inelastic collisions and translate the geometric acceptance of the MBTS into a lower bound of the $\xi$ values probed by the measurement.  The variable $\xi$ is defined at particle level by dividing all final state particles into two systems separated by the largest rapidity gap between two adjacent particles ordered in $\eta$.  The larger mass of the two systems is termed $X$ and $\xi$ is given by $M_{X}^2/s$. There is a strong correlation between $\xi$ and the $\eta$ of the particle furthest from the initial-state proton in $\eta$, leading to a natural translation between $\eta$ acceptance and $\xi$.  There is no restriction on $M_Y$.  

In order to determine the acceptance of the measurement in the restricted $\xi$-range, several models of the diffractive mass spectrum were used: a model from Schuler and Sj\"ostrand~\cite{bib:ssj1} with a relatively flat dependence on $\xi$, a {\sc Phojet} model~\cite{bib:phojetxsec} predicting a slight decrease in the cross-section with decreasing $\xi$, and several power-law based models from Bruni and Ingelman~\cite{bib:bri}, Donnachie and Landshoff (DL)~\cite{bib:dlmodel} and Berger et al.~\cite{bib:berger}. DL predict 
$\frac{{\rm d}\sigma_{SD}}{{\rm d}\xi}  \propto \frac{1}{\xi^{1+\epsilon}}$ where $\epsilon=\alpha(0)-1$ and  $\alpha(t) = \alpha(0) + \alpha^\prime t$.
Values of $\epsilon$ between $0.06$ and $0.10$, and of $\alpha'$ between $0.10$ and $0.40$~GeV$^{-2}$ are considered in this analysis. The DL model with $\epsilon=0.085$ and $\alpha' = 0.25$~GeV$^{-2}$ with {\sc Pythia8} fragmentation is the default model in this analysis and the other models are used to assess uncertainties in the description of diffractive events. The detector response to the generated events is simulated using software~\cite{bib:atlassim} based on {\sc Geant4}~\cite{bib:g4}.

\section{Cross-Section Calculation}
The cross-section is calculated using 
$\sigma_{inel}(\xi>5 \times 10^{-6})~=~\frac{(N-N_{BG})}{\epsilon_{trig}~\times~\int L{\rm d}t}~\times~\frac{1-\ffake}{\epsilon_{sel}}$
where $N$ is the number of selected events, $N_{BG}$ is the number of background events, $\ffake$ is the fraction of events that pass the event selection but have $\xi<5 \times 10^{-6}$, $\int L{\rm d}t$ is the integrated luminosity, and $\epsilon_{trig}$ and $\epsilon_{sel}$ are the trigger and offline event selection efficiencies in the selected $\xi$-range. For $\xi=5\times10^{-6}$, $\epsilon_{sel}$ is 50\%, rising to nearly 100\% for $\xi>10^{-5}$. 

The systematic uncertainties include detector effects such as the MBTS efficiency and detector material distribution, as well as uncertainties on dependence of the correction factors $\epsilon_{sel}$ and $\ffake$ on the underlying diffractive mass spectrum.  The agreement between data and simulation in the detector response was studied by tagging MBTS counters with tracks or calorimeter deposits and comparing the counter charge distributions in data and MC. The fraction of calorimeter tagged counters which registered charged depositions above the hit threshold, a quantity sensitive to the fraction of converted photons, were used to study the material distribution.  The MBTS response contributes 0.1\% to the overall uncertainty on the cross-section and the material distribution contributes 0.2\%.  The trigger efficiency was studied using an independent trigger and found to be known to within 0.1\%.  

The model dependence of $\epsilon_{sel}$ and $\ffake$ was checked using the variety of diffractive mass spectra listed above.  The relative contribution of diffractive to non-diffractive events, $f_D$, was constrained for each model using the fraction of the events which have hits on only one side of the detector in $z$, $R_{SS}$.  Variations within the allowed uncertainties of the relative diffractive contribution contribute 0.3\% to the total uncertainty on the cross-section measurement.  The difference in the correction factors $\epsilon_{sel}$ and $\ffake$ under varying diffractive mass spectra hypothesis leads to a 0.4\% uncertainty.  The  uncertainty due to the fragmentation of the diffractive system was assessed by comparing the correction factors derived with {\sc Pythia 6} and {\sc Pythia 8}, which have significantly different fragmentation mechanisms for diffractive events, and was found to be 0.4\%.  

The background arises from signals in the detector which are not due to proton-proton interactions, such those arising from beam gas, beam halo and cavern radiation.  These sources are estimated to comprise 0.4\% of the events and 100\% uncertainty is assumed. 

The final result for the measured inelastic cross-section is calculated using the default DL model of $\epsilon =0.085$ and $\alpha'=0.25$, 
which yields $f_D=26.9\%$, $\epsilon_{sel}=98.77\%$, and $\ffake=0.96\%$. Together with $\epsilon_{trig}=99.98\%$, $N=1,220,743$, $N_{BG}=1,574$ and $\int L{\rm d}t=20.25$~$\mu$b$^{-1}$ this results in 
$\sigma_{inel}(\xi > 5 \times 10^{-6}) = 60.3 \pm 0.05 {\rm (stat.)} \pm 0.5 {\rm (syst.)} \pm 2.1 {\rm (lumi.)~mb}.$
The luminosity uncertainty of 3.4\% dominates the errors.  

\begin{figure}[htbp]
\includegraphics[width=0.60\textwidth]{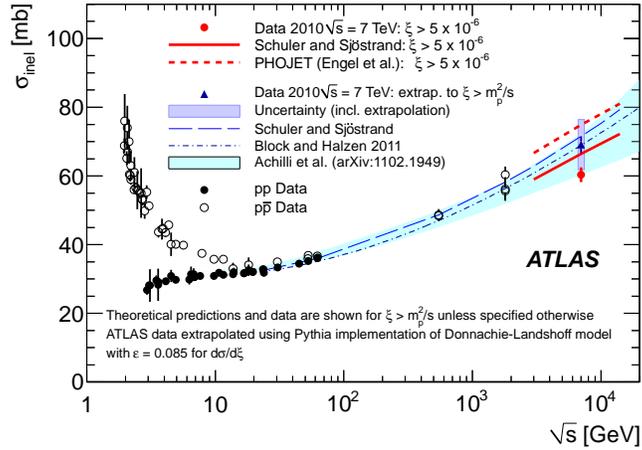}
\caption{ The ATLAS measurement for $\xi > 5 \times 10^{-6}$ (red filled circle) is compared with the Schuler and Sj\"ostrand and {\sc Phojet} models.  Data (filled circles for $pp$, unfilled circles for $p\bar{p}$) from other experiments are compared with predictions of the $pp$ inelastic cross-section. The ATLAS measurement extrapolated to the full inelastic cross-section is also shown (blue triangle).  The error bar indicates the experimental uncertainty while the blue shaded area shows the total (with the extrapolation uncertainty).} 
\label{fig:totxsec}
\end{figure}

 The measurement is compared to the predictions in Figure~\ref{fig:totxsec}. The Schuler-Sj\"ostrand model~(66.4~mb) and the  model~(74.2~mb) predictions are both higher than the data. A prediction of 51.8-56.2~mb by Ryskin~et al.~\cite{bib:kmr2011}, is slightly lower than the data.

To compare with previous measurements and analytic models, the fractional contribution to the inelastic cross-section of events passing the $\xi>5 \times 10^{-6}$ cut is determined from the models and used to extrapolate the measurement to the full inelastic cross-section.  This fraction is 87.3\% for the default model,  and the other models considered give fractions ranging from 96\%~\cite{bib:phojetxsec} to 79\%~\cite{bib:kmr2011}.  Thus 87.3\% is taken as the default value for this fraction and an uncertainty of 10\% is taken due to the extrapolation uncertainty. The resulting inelastic cross-section value is $\sigma_{inel} = 69.1 \pm 2.4 {\rm (exp.)} \pm 6.9 {\rm (extr.)~mb}$ where the experimental uncertainty (exp.) includes the statistical and experimental systematic errors, and the extrapolation (extr.) uncertainty is descibed above.  The extrapolated value agrees within the large uncertainty with the predictions from {\sc Pythia}, which uses a power law dependence on $\sqrt{s}$. It also agrees with Block and Halzen~\cite{bib:block} (which has a logarithmic $\sqrt{s}$ dependence), and with other recent theoretical predictions that vary between 60 and 72~mb~\cite{bib:kmr2011,bib:glm2,bib:godbole}.



\section{Prospects for Elastic Measurements}
In the future, ATLAS will  make complementary measurements of the total proton-proton cross-section, using a far-forward proton tagging detector, ALFA~\cite{bib:alfa}.  Using special LHC runs at high $\beta^{*}$, ALFA will measure the forward elastic scattering cross-section and infer both the total cross-section and luminosity.  ALFA was installed over the 2010-2011 shutdown and high $\beta^{*}$ runs are foreseen for 2011 and 2013/2014.   

\bibliographystyle{aipproc}   


\bibliography{DISProc_Tompkins}



\end{document}